\documentstyle[12pt,subeqn,axodraw]{article}

\newcommand\vek[1]{\mbox{\rmfamily\bfseries\itshape#1}}
\newcommand\vekexp[1]{\mbox{\scriptsize\rmfamily\bfseries\itshape#1}}

\begin{document}

\title{Large-Distance Behavior of Particle Correlations
in the Two-Dimensional Two-Component Plasma}

\author{
L. {\v S}amaj$^{1,2}$ and B. Jancovici$^1$
}

\maketitle

\begin{abstract}
The model under consideration is a two-dimensional 
two-component plasma, i.e., a continuous system of two species 
of pointlike particles of opposite charges $\pm 1$, interacting 
through the logarithmic Coulomb interaction.
Using the exact results for the form-factors of an equivalent
Euclidean sine-Gordon theory, we derive the large-distance behavior
of the pair correlation functions between charged particles.
This asymptotic behavior is checked on a few lower orders of
its $\beta$-expansion ($\beta$ is the inverse temperature)
around the Debye-H\"uckel limit $\beta\to 0$, and at the
free-fermion point $\beta = 2$ at which the collapse of
positive-negative pairs of charges occurs.
\end{abstract}

\medskip

\noindent {\bf KEY WORDS:} Coulomb systems; sine-Gordon model; 
exactly solvable models; correlation functions; form factors.

\bigskip
\noindent LPT Orsay 01-35

\vfill

\noindent $^1$ Laboratoire de Physique Th{\'e}orique, Universit{\'e}
de Paris-Sud, B{\^a}timent 210, 91405 Orsay Cedex, France (Unit{\'e}
Mixte de Recherche no. 8627 - CNRS);

\noindent e-mails: Ladislav.Samaj@th.u-psud.fr 
and Bernard.Jancovici@th.u-psud.fr

\noindent $^2$ On leave from the Institute of Physics, Slovak Academy
of Sciences, Bratislava, Slovakia; 

\noindent e-mail: fyzimaes@savba.sk

\newpage

\renewcommand{\theequation}{1.\arabic{equation}}
\setcounter{equation}{0}

\section{Introduction}
We consider a two-dimensional (2D) two-component plasma (TCP), 
i.e., a neutral continuous system of two species of pointlike 
particles of opposite charges $\pm 1$,
interacting through the 2D logarithmic Coulomb interaction.
Classical equilibrium statistical mechanics is used.
The system is stable against collapse of positive-negative
pairs of charges for the dimensionless coupling constant
(inverse temperature) $\beta < 2$.
In the region $\beta \ge 2$, one has to attach to particles
a hard core in order to prevent the collapse.
In the limit of a small, but nonzero, hard core, the Kosterlitz-Thouless
phase transition from a high-temperature conducting phase
to a low-temperature insulating phase \cite{Kosterlitz}
takes place at around $\beta = 4$.

The 2D TCP is the first continuous fluid in dimension higher than one
with an exactly solvable thermodynamics, in the bulk as well as
at the surface for specific boundary conditions.
The bulk equation of state has been known for a very long time
\cite{Salzberg}.
The other bulk thermodynamic properties (free energy, internal energy,
specific heat, etc.) have been obtained in the whole stability 
range $\beta < 2$ of the plasma in a recent paper \cite{Samaj1}.
The mapping onto a bulk 2D Euclidean (classical) sine-Gordon field 
theory with a specific (conformal) normalization of the cos-field 
was made, and recent results about that field theory 
\cite{Destri} - \cite{Lukyanov1}, were used.
In subsequent works, the surface tension of
the same model in contact with an ideal conductor \cite{Samaj2}
and an ideal dielectric \cite{Samaj3} rectilinear walls was
obtained.
A mapping onto a boundary 2D sine-Gordon field theory with
a Dirichlet and a Neumann boundary condition, respectively,
was made, and known results \cite{Skorik} about those integrable
boundary field theories \cite{Ghoshal} were applied.

In this paper, we derive the large-distance behavior of 
the bulk particle correlation functions for the 2D TCP by exploring
the form-factor theory of the equivalent sine-Gordon model.

Exact sum rules are known for the correlation functions.
The zeroth and second moments of the charge correlation function
have the well-known Stillinger-Lovett values \cite{Stillinger}.
The zeroth moment of the density correlation function is
related to the compressibility (here exactly known), and the
second moment has been recently obtained 
\cite{Jancovici1,Jancovici2}.
An exact analysis of pair correlations in the 2D TCP can be done 
in two cases: in the high-temperature 
Debye-H\"uckel limit $\beta \to 0$ and just at the collapse point 
$\beta = 2$ which corresponds to the free-fermion point of an 
equivalent Thirring model (although, at a given fugacity,
the free energy diverges, truncated particle distributions are 
finite at $\beta=2$ and, as is generally believed, 
also in the whole range $\beta<4$).
We have not found in the literature an exhaustive treatment
of the limit $\beta\to 0$, so we study this limit in detail  
in the present work by using the technique of a renormalized
Mayer expansion in density \cite{Jancovici2,Deutsch}.
At $\beta=2$, the exact forms of the bulk truncated distributions
(Ursell functions) were found in ref. \cite{Cornu1} via 
a continualization of the Gaudin's model \cite{Gaudin}. 
The Ursell functions were computed at $\beta=2$ also for 
inhomogeneous situations when the 2D TCP, being confined to
domains of various geometry, is in contact with a charged 
hard wall, polarizable interface \cite{Cornu2}, an ideal conductor wall 
\cite{Cornu2} - \cite{Jancovici3} and an ideal dielectric wall
\cite{Jancovici4,Tellez}.
A short-distance expansion of pair correlations can be done
in principle for an arbitrary $\beta$ by using Dotsenko's integrals
\cite{Dotsenko}; interestingly, correlations between equally charged 
particles change their analytic structure at short distance at $\beta=1$
\cite{Hansen}.

In the 2D sine-Gordon theory, a differential equation for generalized
correlation functions was derived at the free-fermion point
\cite{Bernard}.
The classical 2D sine-Gordon model can be regarded as a conformal
(namely Gaussian) field theory, perturbed by the cos-field.
For such theories, the short-distance expansion for multipoint
correlation functions can be systematically obtained by using 
the Operator product expansion \cite{Wilson}, combined
with the Conformal perturbation theory \cite{Fateev}.
The most efficient way to study the large-distance behavior
of correlations is provided by the form-factor approach.
For the 2D sine-Gordon model, the form factors of some local
operators for all kinds of particles the theory contains
were evaluated in a series of papers \cite{Karowski} - \cite{Lukyanov3}.
In particular, the form factors for the lightest particle
(elementary breather), dominant when this particle exists,
are analytic continuations of their counterparts
in the sinh-Gordon theory \cite{Fring,Brazhnikov}.
In the sinh-Gordon model, the corresponding particle is the
only massive one; the contributions of all form factors were 
summed up and a closed expression for any correlation function 
as a determinant of an integral operator was obtained in ref. 
\cite{Korepin}.

The paper is organized as follows.
In section 2, we introduce the notation and briefly review 
some important information on the mapping of the 2D TCP onto 
the 2D sine-Gordon model \cite{Samaj1}.
Section 3 deals with the form-factor theory for the 
sine-Gordon model, written in a way accessible to
non-specialists as well.
The large-distance asymptotic of pair particle distributions
for the 2D TCP is presented.
The obtained results are checked around the Debye-H\"uckel
limit $\beta \to 0$ and at the free fermion $\beta=2$ point
in section 4.
A brief recapitulation and some concluding remarks 
are given in section 5.

\renewcommand{\theequation}{2.\arabic{equation}}
\setcounter{equation}{0}

\section{Sine-Gordon representation of the plasma}
We consider an infinite 2D space of points ${\vek r}\in R^2$.
The 2D TCP, defined in this space, consists of point particles
$\{ j \}$ of charge $\{ \sigma_j = \pm 1 \}$, immersed in
a homogeneous medium of dielectric constant $= 1$.
The interaction energy of particles is given by
$\sum_{i<j} \sigma_i \sigma_j v(\vert {\vek r}_i - {\vek r}_j
\vert)$, where the Coulomb potential $v$ is the solution
of the 2D Poisson equation
\begin{equation} \label{2.1}
\Delta v({\vek r}) = - 2 \pi \delta({\vek r})
\end{equation}
Explicitly, $v({\vek r}) = - \ln ( \vert {\vek r} \vert / r_0)$
where $r_0$ is a length scale set for simplicity to unity.
We will work in the grand canonical ensemble, characterized
by the inverse temperature $\beta$ and the equivalent
fugacities of the positively and negatively charged particles,
$z_+ = z_- = z$.
Due to the charge $\pm$ symmetry, the induced particle densities
are $n_+ = n_- = n/2$ ($n$ is the total number density of particles).

Using the fact that $-\Delta/(2\pi)$ is the inverse operator
of $v({\vek r})$ [see equation (\ref{2.1})], the grand partition
function $\Xi$ of the 2D TCP can be turned via the
Hubbard-Stratonovitch transformation
(see, e.g., review \cite{Minnhagen}) into
\begin{equation} \label{2.2}
\Xi(z) = { \int {\cal D} \phi ~ \exp \left( - {\cal A}(z) \right)
\over \int {\cal D} \phi ~ \exp \left( - {\cal A}(0) \right)}
\end{equation}
where 
\begin{subequations} \label{2.3}
\begin{eqnarray}
{\cal A}(z) & = & \int {\rm d}^2 r ~
\left[ {1\over 16 \pi} \left( \nabla \phi \right)^2
-2 z \cos (b \phi) \right] \label{2.3a} \\
b^2 & = & \beta / 4 \label{2.3b}
\end{eqnarray}
\end{subequations}
is the Euclidean action of the 2D classical sine-Gordon theory.
Here, $\phi({\vek r})$ is a real scalar field, $\int {\cal D} \phi$
denotes the functional integration over this field and the
fugacity $z$ is renormalized by a self-energy term.
The sine-Gordon representation of the density of particles 
of one sign is
\begin{eqnarray} 
n_{\sigma} & = & \langle \sum_j \delta_{\sigma,\sigma_j}
\delta({\vek r}-{\vek r}_j) \rangle \nonumber \\
& = & z_{\sigma} \langle {\rm e}^{{\rm i}\sigma b\phi} \rangle
\label{2.4}
\end{eqnarray}
where $\langle \cdots \rangle$ denotes the averaging over
the sine-Gordon action (\ref{2.3}).
The equality of densities $n_+ = n_-$ for the considered
$z_+ = z_- = z$ is a special case of a general symmetry
relation $\langle {\rm e}^{{\rm i}a\phi} \rangle =
\langle {\rm e}^{-{\rm i}a\phi} \rangle$, $a$ arbitrary,
which results from the invariance of the sine-Gordon action
(\ref{2.3}) with respect to the transformation $\phi \to -\phi$.
For two-body densities, one gets
\begin{eqnarray} 
n_{\sigma\sigma'}({\vek r},{\vek r}') & = & \langle 
\sum_{j\ne k} \delta_{\sigma,\sigma_j} \delta({\vek r}-{\vek r}_j) 
\delta_{\sigma',\sigma_k} \delta({\vek r}'-{\vek r}_k) \rangle 
\nonumber \\
& = & z_{\sigma} z_{\sigma'} 
\langle {\rm e}^{{\rm i}\sigma b\phi({\vekexp r})} 
{\rm e}^{{\rm i}\sigma' b\phi({\vekexp r}')} \rangle \label{2.5}
\end{eqnarray}
Clearly, $n_{++}({\vek r},{\vek r}') = n_{--}({\vek r},{\vek r}')$ 
and $n_{+-}({\vek r},{\vek r}') = n_{-+}({\vek r},{\vek r}')$.
For our purpose, it is useful to consider at the two-particle
level also the pair distribution functions
\begin{equation} \label{2.6}
g_{\sigma\sigma'}({\vek r},{\vek r}') =
{n_{\sigma\sigma'}({\vek r},{\vek r}') \over
n_{\sigma} n_{\sigma'}}
\end{equation}
the pair correlation functions
\begin{equation} \label{2.7}
h_{\sigma\sigma'}({\vek r},{\vek r}') =
g_{\sigma\sigma'}({\vek r},{\vek r}') - 1
\end{equation}
and the Ursell functions
\begin{equation} \label{2.8}
U_{\sigma\sigma'}({\vek r},{\vek r}') = n_{\sigma} n_{\sigma'}
h_{\sigma\sigma'}({\vek r},{\vek r}')
\end{equation}
(denoted as $\rho_{\sigma\sigma'}^{(2)T}$ in refs. \cite{Cornu1,Cornu2}).

The parameter $z$ in (\ref{2.3}), i.e., the fugacity renormalized
by a diverging self-energy term, gets a precise meaning when
one fixes the normalization of the cos-field.
In particular, it was shown for the 2D TCP \cite{Hansen},
that the behavior of the two-body densities for oppositely charged 
particles is dominated at short distance by the Boltzmann factor
of the Coulomb potential,
\begin{equation} \label{2.9}
n_{+-}({\vek r},{\vek r}') \sim z_+ z_- \vert {\vek r} - {\vek r}'
\vert^{-\beta} \quad \quad {\rm as}\ 
\vert {\vek r} - {\vek r}' \vert \to 0 
\end{equation}
With respect to (2.5) and the definition (\ref{2.3b}),
one has in the sine-Gordon picture
\begin{equation} \label{2.10}
\langle {\rm e}^{{\rm i}b\phi({\vekexp r})} 
{\rm e}^{-{\rm i}b\phi({\vekexp r}')} \rangle
\sim \vert {\vek r} - {\vek r}' \vert^{-4 b^2} \quad \quad {\rm as}\ 
\vert {\vek r} - {\vek r}' \vert \to 0  
\end{equation}
This short-distance normalization of the exponential fields
makes the mapping complete.
In quantum field theory, (\ref{2.10}) is known as the
conformal normalization, and it corresponds to taking
(\ref{2.3}) as a Gaussian conformal field theory perturbed
by the relevant operator $\cos(b\phi)$.
The conformal normalization makes the bridge between the
2D TCP and the sine-Gordon model, and allows one to transfer
results from one theory to the other \cite{Samaj1}.

Let us now summarize the known facts about the sine-Gordon
theory (\ref{2.3}).
The theory has a discrete symmetry $\phi \to \phi + 2\pi j / b$
($j$ any integer).
This symmetry is spontaneously broken, and the sine-Gordon theory
is massive, in the region $0 < b^2 < 1$ (i.e., $0 < \beta < 4$
for the 2D TCP).
In this region, one has to consider one of infinitely many
ground states $\{ \vert 0_j \rangle \}$ characterized by
$\langle \phi \rangle_j = 2\pi j/b$, say that with $j=0$.
The 2D classical sine-Gordon theory is integrable \cite{Zamolodchikov2}.
Its particle spectrum consists of one soliton-antisoliton pair
$(A_+,A_-)$ (the corresponding particle index $\epsilon = \pm$
denotes the topological $U(1)$ charge) with equal soliton and 
antisoliton masses $m_+=m_-=M$ and of soliton-antisoliton bound 
states (``breathers'') $\{ B_j ; j = 1, 2, \ldots < 1/\xi \}$
(the corresponding particle index $\epsilon = j$), whose number 
at a given $b^2$ depends on the inverse of the parameter
\begin{equation} \label{2.11}
\xi = {b^2 \over 1 - b^2} \quad \quad 
\left( = {\beta \over 4-\beta} \right)
\end{equation}
The range $0<\xi<1$ $(0<\beta<2)$ is an attractive one (breathers exist),
the range $1\le \xi<\infty$ $(2\le \beta<4)$ is a repulsive one 
(breathers do not exist); $\xi=1$ $(\beta=2)$ corresponds to 
the free-fermion (collapse) point.
The masses of breathers $\{ B_j \}$ are given by the formula
\begin{equation} \label{2.12}
m_j = 2 M \sin \left( {\pi \xi \over 2} j \right)
\end{equation}
The lightest breather $B_1$ is usually called the elementary one.
The energy $E$ and the momentum $p$ of a particle $\epsilon$ of mass
$m_{\epsilon}$ are parametrized by the rapidity 
$\theta \in ( -\infty, \infty )$ as follows
\begin{equation} \label{2.13}
E = m_{\epsilon} \cosh \theta , \quad \quad p = m_{\epsilon} \sinh \theta
\end{equation}
Like in other 2D integrable field theories, the $N$-particle
scattering amplitudes are purely elastic (the number of particles
is conserved, the incoming and outgoing momenta are the same) 
and factorized into $N(N-1)/2$ two-particle $S$-matrices which
are determined exactly by exploring their general properties: 
unitarity and crossing symmetry, validity of the Yang-Baxter equation 
and the assumption of ``maximal analyticity''.

The dimensionless specific grand potential $\omega$ of the sine-Gordon
theory (\ref{2.2}), (\ref{2.3}), defined by
\begin{equation} \label{2.14}
- \omega = \lim_{V\to\infty} {1\over V} \ln \Xi
\end{equation}
was found in ref. \cite{Destri} by using the Thermodynamic Bethe
ansatz:
\begin{equation} \label{2.15}
- \omega = {m_1^2 \over 8 \sin(\pi \xi)}
\end{equation}
Under the conformal normalization of the exponential fields
(\ref{2.10}), the relationship between the parameter $z$ and
the soliton mass $M$ was established in ref. \cite{Zamolodchikov1}:
\begin{equation} \label{2.16}
z = {\Gamma(b^2) \over \pi \Gamma(1-b^2)} \left[ M
{\sqrt{\pi} \Gamma((1+\xi)/2) \over 2 \Gamma(\xi/2)}
\right]^{2-2b^2}
\end{equation}
where $\Gamma$ stands for the Gamma function.
As a result,
\begin{equation} \label{2.17}
\langle {\rm e}^{{\rm i}b\phi} \rangle = {1\over 2} 
{\partial (-\omega) \over \partial z} = 
{M^2 \over 8 z (1-b^2)} \tan \left( {\pi \xi \over 2} \right)
\end{equation}
Relations (\ref{2.14}) - (\ref{2.17}), together with the equality
\begin{equation} \label{2.18}
n = 2 z \langle {\rm e}^{{\rm i}b\phi} \rangle
\end{equation}
constitute the complete set of equations for determining
the thermodynamics of the 2D TCP \cite{Samaj1}.
They will be extensively used in the following section.

\renewcommand{\theequation}{3.\arabic{equation}}
\setcounter{equation}{0}

\section{Form-factor theory for the 2D TCP}
In 2D integrable systems, correlation functions of local
operators ${\cal O}_a$ ($a$ is a free parameter) can be
written as an infinite convergent series over multi-particle 
intermediate states.
For the two-point correlation function, one has
\begin{eqnarray}
\langle {\cal O}_a({\vek r}) {\cal O}_{a'}({\vek r}') \rangle
& = & \sum_{N=0}^{\infty} {1\over N!} \sum_{\epsilon_1,\ldots
\epsilon_N} \int_{-\infty}^{\infty} {{\rm d}\theta_1 \ldots
{\rm d}\theta_N \over (2\pi)^N}
F_a(\theta_1,\ldots,\theta_N)_{\epsilon_1\ldots\epsilon_N}
\nonumber \\ & & \times ~
^{\epsilon_N\ldots\epsilon_1}F_{a'}(\theta_N,\ldots,\theta_1)
\exp \left( - \vert {\vek r} - {\vek r}' \vert \sum_{j=1}^N
m_{\epsilon_j} \cosh \theta_j \right) \label{3.1}
\end{eqnarray}
The first $N=0$ term of the series is nothing but the 
decoupling $\langle {\cal O}_a \rangle \langle {\cal O}_{a'} \rangle$.
The form factors
\begin{subequations} \label{3.2}
\begin{eqnarray}
F_a(\theta_1,\ldots,\theta_N)_{\epsilon_1\ldots\epsilon_N}
& = & \langle 0 \vert {\cal O}_a({\vek 0}) \vert 
Z_{\epsilon_1}(\theta_1),\ldots,Z_{\epsilon_N}(\theta_N) \rangle
\\ 
^{\epsilon_N\ldots \epsilon_1}F_{a'}(\theta_N,\ldots,\theta_1)
& = & \langle Z_{\epsilon_N}(\theta_N),\ldots,Z_{\epsilon_1}(\theta_1) 
\vert {\cal O}_{a'}({\vek 0}) \vert 0 \rangle
\end{eqnarray}
\end{subequations}
are the matrix elements of the operator at the origin, between 
an $N$-particle in-state (being a linear superposition of free
one-particle states $\vert Z_{\epsilon}(\theta) \rangle$)
and the vacuum.
They depend on particle rapidities only through their differences
$\theta_{jk} = \theta_j - \theta_k$.
The form factors can be determined exactly (for the underlying 
sine-Gordon theory, see refs. \cite{Karowski}-\cite{Lukyanov3}).
They satisfy a set of constraint functional equations,
originating from general properties of unitarity, analyticity,
relativistic invariance and locality.
The important fact is that these functional equations do not
refer to a specific operator ${\cal O}$, the $S$-matrix
is the only dynamical information needed.
The nature of the operator is reflected only via normalization
constants for the form factors.
In what follows, under ${\cal O}_a$ we will understand an
exponential operator
\begin{equation} \label{3.3}
{\cal O}_a({\vek r}) = \exp \left( {\rm i} a \phi({\vek r}) \right)
\end{equation}

The form-factor representation of two-point correlation functions
(\ref{3.1}) is particularly useful for large distance
$\vert {\vek r}-{\vek r}' \vert$.
In the limit $\vert {\vek r}-{\vek r}' \vert \to \infty$, the
dominant contribution to the truncated function
$\langle {\cal O}_a({\vek r}) {\cal O}_{a'}({\vek r}') \rangle_{{\rm T}}
= \langle {\cal O}_a({\vek r}) {\cal O}_{a'}({\vek r}') \rangle
- \langle {\cal O}_a \rangle \langle {\cal O}_{a'} \rangle$
comes from a multi-particle intermediate state with the
minimum value of the total particle mass $\sum_{j=1}^N m_{\epsilon_j}$,
at the point of vanishing rapidities $\theta_j \to 0$.
The corresponding exponential decay 
$\exp(-\vert {\vek r}-{\vek r}' \vert \sum_{j=1}^N m_{\epsilon_j})$
is modified by a slower (inverse power law) decaying function
which particular form depends on the form factors.
For the sine-Gordon theory (\ref{2.3}) (or, equivalently,
the 2D TCP), in the region $0 < b^2 <1/2$ $(0 < \beta < 2)$,
the lightest particle in the spectrum is the elementary breather $B_1$.
As is clear from (\ref{2.12}), approaching $b^2 \to 1/2$
resp. $\xi \to 1$ $(\beta \to 2)$, its mass $m_1 \to 2M$.
Note that, according to eq. (\ref{2.16}), $M$ is finite
at $b^2 = 1/2$, resp. $\xi = 1$ for a fixed value of $z$.
The elementary breather $B_1$ disappears at the free-fermion point 
$b^2 = 1/2$ (collapse point $\beta = 2$).
The only existing particles remain the topologically neutral 
soliton-antisoliton pair $(A_+,A_-)$, which has just the same 
total mass $2M$.
Thus, the total mass remains $2M$ (which is temperature-dependent)
in the whole region $1/2 \le b^2 < 1$ $(2 \le \beta < 4)$.

\subsection{ $0 < b^2 < 1/2$ $(0 < \beta <2)$}
For the elementary breather $B_1$, the form factors
$F_a(\theta_1,\ldots,\theta_N)_{1\ldots 1}$ and
$^{1\ldots 1}F_{a'}(\theta_N,\ldots,\theta_1)$ 
$=$ $F_{a'}(\theta_N,\ldots,\theta_1)_{1\ldots 1}$
are presented for the exponential operator (\ref{3.3})
in ref. \cite{Lukyanov3}.
In the notation (\ref{3.2}), they read
\begin{subequations} \label{3.4}
\begin{eqnarray}
\langle 0 \vert {\rm e}^{{\rm i}a\phi} \vert B_1(\theta) \rangle
& = & - {\rm i} ~ \langle {\rm e}^{{\rm i}a\phi} \rangle
~ (\pi \lambda)^{1/2} ~ {\sin(\pi\xi a/b) \over \sin(\pi\xi)}
\label{3.4a} \\
\langle 0 \vert {\rm e}^{{\rm i}a\phi} \vert B_1(\theta_2),
B_1(\theta_1) \rangle 
& = & - \langle {\rm e}^{{\rm i}a\phi} \rangle
~ (\pi \lambda) ~ \left[ {\sin(\pi\xi a/b) \over \sin(\pi\xi)}
\right]^2 R(\theta_1 - \theta_2) \label{3.4b}
\end{eqnarray}
\end{subequations} 
etc., where
\begin{equation} \label{3.5}
\lambda = {4\over \pi} \sin(\pi\xi) \cos \left( {\pi \xi \over 2}
\right) \exp \left\{ - \int_0^{\pi\xi} {{\rm d}t \over \pi}
{t\over \sin t} \right\} 
\end{equation}
and the function $R(\theta)$ is represented in the range 
$-2\pi + \pi\xi < {\rm Im}(\theta) < -\pi\xi$ by the integral
\begin{subequations}
\begin{eqnarray} 
R(\theta) & = & {\cal N} \exp \left\{ 8 \int_0^{\infty}
{{\rm d}t \over t} { \sinh t \sinh(t\xi) \sinh(t(1+\xi)) 
\over \sinh^2(2t)} \sinh^2 t\left( 1 - 
{{\rm i}\theta \over \pi} \right) \right\} \label{3.6a} \\
{\cal N} & = & \exp \left\{ 4 \int_0^{\infty}
{{\rm d}t \over t} { \sinh t \sinh(t\xi) \sinh(t(1+\xi)) 
\over \sinh^2(2t)} \right\} \label{3.6b}
\end{eqnarray}
\end{subequations}
An analytic continuation of this representation
to the case of interest ${\rm Im}(\theta) = 0$ is given by
the relation
\begin{equation} \label{3.7}
R(\theta) R(\theta \pm {\rm i}\pi) =
{\sinh \theta \over \sinh \theta \mp {\rm i} \sin(\pi\xi)}
\end{equation}
Note a different notation in comparison with ref. \cite{Lukyanov3}.

The pair correlation function $h_{\sigma\sigma'}({\vek r},{\vek r}')$ 
(\ref{2.7}) is expressible as follows
\begin{equation} \label{3.8}
h_{\sigma\sigma'}({\vek r},{\vek r}') =
{ \langle {\rm e}^{{\rm i}\sigma b\phi({\vekexp r})}
{\rm e}^{{\rm i}\sigma' b\phi({\vekexp r}')} \rangle_{{\rm T}}
\over \langle {\rm e}^{{\rm i}\sigma b\phi} \rangle 
\langle {\rm e}^{{\rm i}\sigma' b\phi} \rangle }
\end{equation}
Using the form-factor representation (\ref{3.1}),
the dominant contribution to $h_{\sigma\sigma'}({\vek r},{\vek r}')$ 
in the limit $\vert{\vek r}-{\vek r}'\vert \to \infty$
is given by the one-breather $B_1$ state with mass $m_1$
and the form factor (\ref{3.4a}).
Then, since
\begin{equation} \label{3.9}
\int_{-\infty}^{\infty} {{\rm d}\theta \over 2}
{\rm e}^{-r m_1 \cosh\theta} =  K_0(m_1 r)
\sim \left( { \pi \over 2 m_1 r}\right)^{1/2} \exp(- m_1 r)
\end{equation}
where $K_0$ is the modified Bessel function of second kind
\cite{Gradshteyn}, one finds, at asymptotically large distance, that
\begin{equation} \label{3.10}
h_{\sigma \sigma'}(r) \sim \sigma \sigma' h(r)
\quad \quad \quad {\rm as}\ r\to\infty
\end{equation}
with
\begin{equation} \label{3.11}
h(r) = - \lambda \left( {\pi \over 2 m_1 r}\right)^{1/2}
\exp( - m_1 r )
\end{equation}
Using the thermodynamic formulae derived in section 2, 
the lightest-breather mass $m_1$ is expressed as follows 
\begin{subequations} \label{3.12}
\begin{eqnarray}
m_1 & = & \kappa \left[ { \sin(\pi\beta/(4-\beta)) \over
\pi\beta/(4-\beta)} \right]^{1/2} \label{3.12a} \\
& = & \kappa \left[ 1 - {\pi^2 \over 192} \beta^2
- {\pi^2 \over 384} \beta^3 + O(\beta^4) \right] \label{3.12b}
\end{eqnarray}
\end{subequations}
where
\begin{equation} \label{3.13}
\kappa = (2\pi\beta n)^{1/2}
\end{equation}
denotes the inverse Debye length.
The parameter $\lambda$ (\ref{3.5}) takes the form
\begin{subequations} \label{3.14}
\begin{eqnarray}
\lambda & = & {4\over \pi} \sin\left( {\pi \beta \over 4-\beta} \right)
\cos\left( {\pi \beta \over 2(4-\beta)} \right)
\exp\left\{ - \int_0^{\pi\beta \over 4-\beta} {{\rm d}t \over \pi}
{t\over \sin t} \right\} \label{3.14a} \\
& = & \beta \left[ 1 - \left( {1\over 32}+{7\pi^2\over 384}
\right) \beta^2 - \left( {1\over 96}+{23\pi^2 \over 2304} \right)
\beta^3 + O(\beta^4) \right] \label{3.14b}
\end{eqnarray}
\end{subequations}

The specific dependence of $h_{\sigma\sigma'}(r)$ on the
product of charges $\sigma\sigma'$, formula (\ref{3.10}),
means that the two-particle correlations are determined
at large distances by the charge-charge correlation function.
Indeed,
\begin{equation} \label{3.15}
{1\over 4} \sum_{\sigma,\sigma'=\pm} \sigma \sigma'
h_{\sigma\sigma'}(r) = h(r)
\end{equation}
On the other hand, the density-density correlation function
$\sum_{\sigma,\sigma'=\pm} h_{\sigma\sigma'}(r)/4$ 
vanishes for the lowest one-breather
$B_1$ state, and becomes nonzero only for the next 
two-breather $B_1$ state, with the form factor (\ref{3.4b})
and much faster exponential decay $\exp(-2 m_1 r)$.

\subsection{ $1/2 \le b^2 < 1$ $(2 \le \beta < 4)$ }
The form factors for the soliton-antisoliton pair with
topological $U(1)$-charges $\epsilon = \pm$ are nonvanishing
only for $U(1)$ neutral states $\sum_{j=1}^N \epsilon_j = 0$,
with $N$ being inevitably an even number.
Simultaneously, 
$^{\epsilon_N\ldots\epsilon_1}F_{a'}(\theta_N,\ldots,\theta_1)
= F_{a'}(\theta_N,\ldots,\theta_1)_{\epsilon_N^*\ldots\epsilon_1^*}$
where $\epsilon^* = - \epsilon$.
Form factors with different assignments of the charges 
$\{ \epsilon_i \}$ are related by the relation \cite{Bernard}
\begin{equation} \label{3.16}
F_a(\theta_1,\ldots,\theta_i,\theta_{i+1},\ldots
\theta_N)_{\epsilon_1\ldots\epsilon_i\epsilon_{i+1}\ldots\epsilon_N} =
- F_a(\theta_1,\ldots,\theta_{i+1},\theta_i,\ldots
\theta_N)_{\epsilon_1\ldots\epsilon_{i+1}\epsilon_i\ldots\epsilon_N}
\end{equation}
and the hermiticity of the operator (\ref{3.3}) implies
\begin{equation} \label{3.17}
F_a(\theta_1,\ldots,\theta_N)_{\epsilon_1\ldots\epsilon_N} =
F_{-a}(\theta_N,\ldots,\theta_1)_{\epsilon_N^*\ldots\epsilon_1^*}
\end{equation}
In the notation (\ref{3.2}), the two-particle form factors for
the exponential operator (\ref{3.3}) with the special value
of $a=b$ read \cite{Lukyanov3}
\begin{equation} \label{3.18}
\langle 0 \vert {\rm e}^{{\rm i}b\phi} \vert A_{\pm}(\theta_2),
A_{\mp}(\theta_1) \rangle = \langle {\rm e}^{{\rm i}b\phi} \rangle
G_{\mp\pm}^b(\theta_1-\theta_2)
\end{equation}
where
\begin{subequations} \label{3.19}
\begin{eqnarray}
G_{\mp\pm}^b(\theta) & = & - 2 G(\theta) \sinh(\theta) 
{\rm cotg}\left( {\pi \xi \over 2} \right) \exp\left( \mp
{\theta +{\rm i}\pi \over 2 \xi} \right) \left[ \xi 
\sinh\left( {\theta + {\rm i}\pi \over \xi} \right) \right]^{-1}
\label{3.19a} \\
G(\theta) & = & \exp \left\{ \int_0^{\infty} {{\rm d}t \over t}
{\sinh(t(\xi-1)) \over \sinh(2t) \cosh t \sinh(t\xi)}
\sinh^2 t\left( 1-{{\rm i}\theta \over \pi} \right) \right\}
\label{3.19b}
\end{eqnarray}
\end{subequations}
Note a different notation.
The form factors of the operator $\exp(-{\rm i}b\phi)$ are
deducible from (\ref{3.18}) and (\ref{3.19}) by using 
relation (\ref{3.17}). 

In the region $2\le \beta <4$, for fixed $z$, the total particle 
density $n$ is infinite, so that only the fugacity $z$ can be chosen
as a legitimate parameter.
Instead of $h$, which vanishes in this region, the Ursell function
(\ref{2.8}) is considered.
Using (\ref{3.16})-(\ref{3.19}) in the form-factor representation
(\ref{3.1}) given by the contribution of the soliton-antisoliton
pair, we get, as previously,
\begin{equation} \label{3.20}
U_{\sigma \sigma'}(r) \sim \sigma \sigma' U(r)
\quad \quad \quad {\rm as}\ r\to\infty
\end{equation}
where $U(r)$ is the leading asymptotic $r\to \infty$
term of the integral
\begin{eqnarray} \label{3.21}
& & - {M^4 \cos(\pi/\xi) \over 16 b^4} \int_{-\infty}^{\infty}
{{\rm d}\theta_1 {\rm d}\theta_2 \over (2\pi)^2}
G(\theta) G(-\theta) \sinh^2\theta \nonumber \\
& & \quad \quad \quad \quad \quad \quad \times \left[ 
\sinh \left( {\theta + {\rm i}\pi \over \xi}\right)
\sinh \left( {-\theta + {\rm i}\pi \over \xi}\right) \right]^{-1}
{\rm e}^{-M r (\cosh \theta_1 + \cosh \theta_2)} 
\end{eqnarray}
with $\theta = \theta_1 - \theta_2$.

At $\beta=2$, since $\xi=1$ it holds $G(\theta)=1$, and therefore
\begin{subequations} \label{3.22}
\begin{equation} \label{3.22a}
U(r) = - {m^3 \over 8\pi r} \exp ( - 2 m r) \ ,
\quad \quad \quad \beta = 2
\end{equation}
where
\begin{equation} \label{3.22b}
m = 2 \pi z
\end{equation}
\end{subequations}
is the soliton mass $M$ at $\beta=2$, see formula (\ref{2.16})
with $b^2=1/2$ and $\xi=1$.

For $\beta>2$, it can be readily shown from (\ref{3.21}) that
\begin{subequations} \label{3.23}
\begin{equation} \label{3.23a}
U(r) \propto - {1 \over r^2} \exp ( - 2 M r) \ ,
\quad \quad \quad 2< \beta <4
\end{equation}
where the soliton mass $M$ is given by formula (\ref{2.16})
in the following form
\begin{equation} \label{3.23b}
M = {2 \Gamma(\beta/[2(4-\beta)]) \over \sqrt{\pi}
\Gamma(2/(4-\beta))} \left[ {\pi z \Gamma(1-\beta/4)
\over \Gamma(\beta/4)} \right]^{1\over 2-\beta/2}
\end{equation}
\end{subequations}
The prefactor in (\ref{3.23a}) could, in principle, be obtained 
from (\ref{3.21}), but we omit this complicated calculation.

\renewcommand{\theequation}{4.\arabic{equation}}
\setcounter{equation}{0}

\section{Analytic checks of the results}

\subsection{Small coupling expansion}
In this section, the small-$\beta$ expansion (\ref{3.12}) and 
(\ref{3.14}) of the asymptotic form (\ref{3.11}) of the charge 
correlation function will be checked by a direct
calculation using the renormalized Mayer expansion 
\cite{Samaj1,Jancovici2}.

The two-dimensional Fourier transforms which will be needed are defined
here, for instance for the charge correlation function $h(r)$, as
\begin{subequations}
\begin{equation} \label{4.1a}
\hat{h}(k)=\int_0^{\infty} h(r) J_0(kr) r {\rm d}r
\end{equation}
\begin{equation} \label{4.1b}
h(r)=\int_0^{\infty} \hat{h}(k) J_0(kr) k {\rm d}k
\end{equation}
\end{subequations}
where $J_0$ is a Bessel function. 
$\hat{h}(k)$ is related to the Fourier transform $\hat{c}(k)$ 
of the charge direct correlation function by the Ornstein-Zernike
relation
\begin{equation} \label{4.2}
\hat{h}(k)=\frac{\hat{c}(k)}{1-2\pi n\hat{c}(k)}
\end{equation}
where $n$ is the total number density of the particles (in ref.
\cite{Jancovici2}, $h$ and $c$ were called $h'$ and $c'$, respectively).
  
At the lowest order in $\beta$ (Debye-H\"{u}ckel approximation), 
$c(r)=\beta \ln r$, $\hat{c}(k)=-\beta/k^2$, and
\begin{equation} \label{4.3}
\hat{h}(k)=-\frac{\beta}{k^2+\kappa^2} 
\end{equation}
where $\kappa$ is the inverse Debye length defined in eq. (\ref{3.13}).
Thus,
\begin{equation} \label{4.4} 
h(r)=-\beta K_0(\kappa r)
\end{equation}
where $K_0$ is a modified Bessel function. 
The asymptotic form of $h(r)$ is \cite{Gradshteyn}
\begin{equation} \label{4.5}
h(r) = - \beta \left( {\pi \over 2 \kappa r} \right)^{1/2}
\exp (-\kappa r) 
\end{equation}
in agreement with (\ref{3.11}) - (\ref{3.14}).

Further terms in the $\beta$-expansion of $c(r)$ are given by the 
renormalized Mayer expansion in density.
Graphs of the excess Helmholtz free energy contributing to
$c(r)$ are the ones which have their two root-vertices with
on odd bond-coordination and their field vertices with an even
bond-coordination.  
Up to order $\beta^4$, for a fixed value of $\kappa$, 
\begin{equation} \label{4.6}
c(r) = \beta \ln r + \beta^3 c_3(r) + \beta^4 c_4(r)
\end{equation}
where $\beta^3c_3$ and $\beta^4c_4$ are defined by the diagrams [where 
a wavy line represents the renormalized bond $-\beta K_0(\kappa r)$]:
\begin{subequations}
\begin{equation} \label{4.7a}
\beta^3 c_3(r) = \ \ \ 
\begin{picture}(55,40)(0,7)
    \PhotonArc(20,6)(20,15,165){1}{11}
    \PhotonArc(20,14)(20,195,345){1}{11}
    \Photon(0,10)(40,10){1}{8.5}
    \BCirc(0,10){2.5} \BCirc(40,10){2.5}
\end{picture}
= - \frac{\beta^3}{6} K_0^3(\kappa r)
\end{equation}
and
\begin{equation} \label{4.7b}
\beta^4 c_4(r) = \ \ \ 
\begin{picture}(60,40)(0,16)
    \PhotonArc(32,6)(32,115,170){1}{7.5}
    \PhotonArc(-12,40)(32,295,355){1}{7.5}
    \PhotonArc(9,5)(32,10,70){1}{7.5}
    \PhotonArc(52,40)(32,185,250){1}{7.5}
    \Photon(0,10)(40,10){1}{8}
    \BCirc(0,10){2.5} \BCirc(40,10){2.5} \Vertex(20,35){2.5}
\end{picture}
= - \frac{\beta^5}{4} K_0(\kappa r)
n\kappa^{-2}\int{\rm d}^2(\kappa r')\,K_0^2(\kappa r')
K_0^2(\kappa \vert {\vek r}-{\vek r}' \vert )
\end{equation}
\end{subequations}
\medskip
\noindent 
These diagrams are of order $\beta^3$ and $\beta^4$, respectively 
[this is apparent for $c_4(r)$ when the factor $n\kappa^{-2}$ in front 
of the integral is replaced by $1/(2\pi\beta)$]. 
$c(r)$ has no term of order $\beta^2$. 
The Fourier transforms of $c_3(r)$ and $c_4(r)$ will now be shown 
to have convenient integral representations. 
From now on, $\kappa^{-1}$ will be taken as the unit of length.

In the Fourier transform
\begin{equation} \label{4.8}
\hat{c}_3(k)=-\frac{1}{6}\int_0^{\infty} K_0^3(r)J_0(kr)r{\rm d}r
\end{equation}
the integral can be viewed as $1/(2\pi)$ times the scalar product of 
$K_0^2(r)$ and $K_0(r)J_0(kr)$, which can be written as
the scalar product of their Fourier transforms. These transforms 
are \cite{Samaj1,Gradshteyn}
\begin{equation} \label{4.9}
\hat{G}(l)=\int_0^{\infty} K_0^2(r)J_0(lr) r {\rm d}r =
\frac{\ln \left[{l\over 2} + \sqrt{1+\left({l\over 2}\right)^2}\right]}
{l ~ \sqrt{1+\left(\frac{l}{2}\right)^2}}
\end{equation}
and
\begin{equation} \label{4.10}
\hat{H}_k(l)=\int_0^{\infty} K_0(r)J_0(kr)J_0(lr)r{\rm d}r=
(1+k^4+l^4-2k^2l^2+2k^2+2l^2)^{-1/2}
\end{equation}
Therefore, setting $l=2\sinh\varphi$, one finds
\begin{eqnarray} \label{4.11}
\hat{c}_3(k)&=&-\frac{1}{6}\int_0^{\infty} \hat{G}(l)\hat{H}_k(l)l
{\rm d}l \nonumber \\
&=&-\frac{1}{3}\int_0^{\infty}\frac{{\rm d}\varphi ~ \varphi}
{\left[16\sinh^4\varphi+16\sinh^2\varphi-8(k^2+1)\sinh^2\varphi+
(k^2+1)^2\right]^{1/2}}
\end{eqnarray}
In the expression (\ref{4.7b}) of $c_4(r)$, there is an integral $I(r)$,
which is the convolution of $K_0^2$ with itself. Its Fourier 
transform is $\hat{I}(l)=2\pi[\hat{G}(l)]^2$. Again, in
\begin{equation} \label{4.12}
\hat{c}_4(k)=-\frac{1}{8\pi}\int_0^{\infty}
 K_0(r)I(r)J_0(kr)r{\rm d}r
\end{equation}
the integral can be viewed as $1/(2\pi)$ times a scalar product, now of
$I(r)$ and $K_0(r)J_0(kr)$. As above, in terms of the 
Fourier transforms $\hat{I}(l)$ and $\hat{H}_k(l)$, 
\begin{equation} \label{4.13}
\hat{c}_4(k)=-\frac{1}{8\pi}\int_0^{\infty}\hat{I}(l)\hat{H}_k(l)l{\rm d}l
=-\frac{1}{4}\int_0^{\infty} [\hat{G}(l)]^2 \hat{H}_k(l)l{\rm d}l
\end{equation}
Using (\ref{4.9}) and (\ref{4.10}) in (\ref{4.13}), and again setting
$l=2\sinh\varphi$, one finds
\begin{equation} \label{4.14}
\hat{c}_4(k)=
-\frac{1}{4}\int_0^{\infty}\frac{{\rm d}\varphi ~ \varphi^2}
{\sinh\varphi\cosh\varphi\left[16\sinh^4\varphi+16\sinh^2\varphi
-8(k^2+1)\sinh^2\varphi+(k^2+1)^2\right]^{1/2}} 
\end{equation}
Finally, up to order $\beta^4$,
\begin{equation} \label{4.15}
\hat{c}(k)=-\frac{\beta}{k^2}[1-\beta^2k^2\hat{c}_3(k)-
\beta^3k^2\hat{c}_4(k)]
\end{equation}
where $\hat{c}_3$ and $\hat{c}_4$ are given by (\ref{4.11}) 
and (\ref{4.14}).

Using (\ref{4.15}) in (\ref{4.2}) (with $\kappa=1$) gives
\begin{equation} \label{4.16}
\hat{h}(k)=-\frac{\beta[1-\beta^2k^2\hat{c}_3(k)-\beta^3k^2\hat{c}_4(k)]}
{k^2+1-\beta^2k^2\hat{c}_3(k)-\beta^3k^2\hat{c}_4(k)}
\end{equation}
We are interested in the asymptotic behavior of $h(r)$, which is governed
by the poles of $\hat{h}(k)$ closest to the real axis. 
When $\beta\rightarrow0$, these poles are at $k=\pm {\rm i}$, 
i.e., at $k^2=-1$.
For finding the location and the contributions of these poles 
in a $\beta$ expansion when the $\beta^2$ and $\beta^3$ terms 
in the denominator of (\ref{4.16}) are taken into account, 
it is enough to expand $\hat{c}_3$ and $\hat{c}_4$ 
around $k^2=-1$ up to first order in $1+k^2$. 

However, some care is needed for deriving this
expansion from the expressions (\ref{4.11}) and (\ref{4.14}). 
Indeed, these expressions were obtained by using the Fourier transform 
(\ref{4.10}) of $K_0(r)J_0(kr)$. 
However, for complex $k$, this integral diverges at large $r$ when 
$\vert {\rm Im}\,k \vert>1$. 
Thus, the integral representations (\ref{4.11}) and (\ref{4.14}) 
are valid for $1+k^2$ real positive but are expected to exhibit 
some singularity at $1+k^2=0$ [although the original function 
$\hat{c}_3(k)$ defined by (\ref{4.8}) and its counterpart for 
$\hat{c}_4(k)$ (\ref{4.12}) are regular around $1+k^2=0$; for
instance, (\ref{4.8}) defines a function of $k$ which is analytical in
the whole strip $|{\rm Im}\,k|<3$, thus in particular for any real value
of $1+k^2>-8$]. 
The singularity in (\ref{4.11}) or (\ref{4.14}) can be 
seen directly if a naive expansion with respect to 
$x=1+k^2$ is attempted: the coefficient of $x^2$ is an integral which 
diverges at small $\varphi$. 
This is a warning that one must be careful when computing 
the coefficient of the previous term (of order $x$) in
the expansion, as follows.

The zeros of the denominator in (\ref{4.11}) are $\sinh^2\varphi=
-(1/2)[1-(x/2)]\pm (1-x)^{1/2}]$, i.e., for small $x$, 
\begin{equation} \label{4.17}
\hat{c}_3(k)=-\frac{1}{12}\int_0^{\infty}\frac{{\rm d}\varphi ~
\varphi}
{[\sinh^2\varphi+(x^2/16)+\ldots]^{1/2}[\cosh^2\varphi-(x/2)+\ldots]
^{1/2}}
\end{equation}
While the second factor in the denominator can be expanded with respect 
to $x$, the first factor is the dangerous one in which the $x^2$ term
will give a contribution of first order in $x$ and should not be
suppressed. To first order in $x$,
\begin{eqnarray} \label{4.18} 
\hat{c}_3(k)&=&-\frac{1}{12}\int_0^{\infty}\frac{{\rm d}\varphi ~ 
\varphi}
{\cosh\varphi [\sinh^2\varphi+(x^2/16)]^{1/2}}\left[1+\frac{x}
{4\cosh^2\varphi}\right] \nonumber \\
&=&-\frac{1}{12}\int_0^{\infty}\frac{{\rm d}\varphi ~ \sinh\varphi}
{\cosh\varphi [\sinh^2\varphi+(x^2/16)]^{1/2}} \nonumber \\
&+&\frac{1}{12}\int_0^{\infty}\frac{(\sinh\varphi-\varphi){\rm d}\varphi}
{\cosh\varphi [\sinh^2\varphi+(x^2/16)]^{1/2}} \nonumber \\
&-&\frac{x}{48}\int_0^{\infty}\frac{\varphi{\rm d}\varphi}
{\cosh^3\varphi [\sinh^2\varphi+(x^2/16)]^{1/2}}
\end{eqnarray}
In an expansion of $\hat{c}_3$ to first order in $x$, the third integral
in the rhs of (\ref{4.18}) is needed only for $x^2=0$. By an integration
per partes, it is found to be
\begin{equation} \label{4.19}
\int_0^{\infty}\frac{{\rm d}\varphi ~ \varphi}
{\cosh^3\varphi \sinh\varphi}=\frac{\pi^2}{8}-\frac{1}{2}
\end{equation}
A naive expansion of the second integral in the rhs of (\ref{4.18})
gives a finite coefficient for the $x^2$ term. 
Thus, the integral has no term of first order in $x$ 
and can be evaluated at $x^2=0$. 
Using tabulated integrals \cite{Gradshteyn}, one finds
\begin{equation} \label{4.20}
\int_0^{\infty}\frac{{\rm d}\varphi ~ (\sinh\varphi-\varphi)}
{\cosh\varphi\sinh\varphi}=\frac{\pi}{2}-\frac{\pi^2}{8}
\end{equation}
Finally, the first integral in the rhs of (\ref{4.18}), which is the
delicate one, can be exactly evaluated, by taking $\cosh\varphi$ as the
integration variable. 
The result does have a term of order $x$, which would have been missed 
if $x^2$ had been neglected in the integrand:
\begin{equation} \label{4.21}
\int_0^{\infty}\frac{{\rm d}\varphi ~ \sinh\varphi}
{\cosh\varphi [\sinh^2\varphi+(x^2/16)]^{1/2}}=
\frac{\pi}{2}-\frac{x}{4}+\ldots
\end{equation}
It should be noted that (\ref{4.21}) has been derived for $x$ real
positive. 
For $x$ real negative, $x$ must be replaced by $|x|$ in the
rhs of (\ref{4.21}), which confirms that the integral representation 
(\ref{4.11}) has a singularity (a kink) at $x=0$. 
However, since the original expression (\ref{4.8}) is regular around 
$x=0$ and the integral representation (\ref{4.11}) is valid for $x$ real 
positive, (\ref{4.21}) can be used for obtaining the correct analytical 
behavior of $\hat{c}_3(k)$ near $1+k^2=0$. 
Using (\ref{4.19}), (\ref{4.20}), and 
(\ref{4.21}) in (\ref{4.18}) gives the $1+k^2$ expansion
\begin{equation} \label{4.22}
\hat{c}_3(k)=-\frac{\pi^2}{96}+\left( \frac{1}{32}-\frac{\pi^2}{384}
\right) (1+k^2)+\ldots
\end{equation}

A similar approach can be used for expanding the expression (\ref{4.14})
of $\hat{c}_4(k)$. 
To first order in $x=1+k^2$,
\begin{eqnarray} \label{4.23} 
\hat{c}_4(k)&=&-\frac{1}{16}\int_0^{\infty}\frac{{\rm d}\varphi ~ 
\sinh\varphi}
{\cosh^2\varphi[\sinh^2\varphi+(x^2/16)]^{1/2}} \nonumber \\
&+&\frac{1}{16}\int_0^{\infty}\frac{{\rm d}\varphi ~
(\sinh^2\varphi-\varphi^2)}{\sinh^2\varphi\cosh^2\varphi} \nonumber \\
&-&\frac{x}{64}\int_0^{\infty}\frac{{\rm d}\varphi ~ \varphi^2}
{\sinh^2\varphi\cosh^4\varphi}
\end{eqnarray}
By simple manipulations, the third integral in the rhs of (\ref{4.23}) 
is found to be 
\begin{equation} \label{4.24}
\int_0^{\infty}\frac{{\rm d}\varphi ~ \varphi^2}
{\sinh^2\varphi\cosh^4\varphi}=\frac{\pi^2}{36}+\frac{1}{3}
\end{equation}
From tables \cite{Gradshteyn}, the second integral is
\begin{equation} \label{4.25}
\int_0^{\infty}\frac{{\rm d}\varphi ~ 
(\sinh^2\varphi-\varphi^2)}{\sinh^2\varphi\cosh^2\varphi}
=1-\frac{\pi^2}{12}
\end{equation}
Finally, the first integral can be exactly evaluated and afterwards
expanded in $x$ with the result
\begin{equation} \label{4.26}
\int_0^{\infty}\frac{{\rm d}\varphi ~ \sinh\varphi}
{\cosh^2\varphi[\sinh^2\varphi+(x^2/16)]^{1/2}}
=1-\frac{x}{4}+\ldots
\end{equation}
Using (\ref{4.24}), (\ref{4.25}), and (\ref{4.26}) in (\ref{4.23}) gives
\begin{equation} \label{4.27}
\hat{c}_4(k)=-\frac{\pi^2}{192}+\left(\frac{1}{96}-\frac{\pi^2}
{2304}\right)(1+k^2)+\ldots
\end{equation}

By using (\ref{4.22}) and (\ref{4.27}) in the denominator of 
(\ref{4.16}), it is seen that, at order $\beta^3$, this denominator has 
poles at $k=\pm{\rm i}m_1$ with 
\begin{equation} \label{4.28}
m_1 = 1-\frac{\pi^2}{192}\beta^2 - \frac{\pi^2}{384}\beta^3
\end{equation}
The asymptotic behavior of $h(r)$ is governed by these poles and the
corresponding residues, i.e., by writing $\hat{h}(k)$ in the  
form $-\lambda(k^2)(k^2+m_1^2)^{-1}$ and replacing $k^2$ by $-m_1^2$ 
in $\lambda(k^2)$. 
One obtains  
\begin{equation} \label{4.29}
\hat{h}(k)\sim - \frac{\lambda}{k^2+m_1^2} 
\end{equation}
where, at order $\beta^3$,
\begin{equation} \label{4.30}
\lambda = \beta \left[ 
1-\left(\frac{1}{32}+\frac{7\pi^2}{384}\right)\beta^2-
\left(\frac{1}{96}+\frac{23\pi^2}{2304}\right)\beta^3 \right]
\end{equation}
The corresponding $h(r)$ is $- \lambda K_0(m_1r)$ which has 
the asymptotic behavior (in units of $\kappa^{-1}$) 
\begin{equation} \label{4.31}
h(r)\sim - \lambda \left( {\pi \over 2 m_1 r}\right)^{1/2}
\exp(- m_1 r)
\end{equation}
with $m_1$ and $\lambda$ given by (\ref{4.28}) and (\ref{4.30}). 
This is in agreement with the small-$\beta$ expansions 
of these quantities (\ref{3.12b}) and (\ref{3.14b}) obtained
from the exact result.

\subsection{ Free-fermion $\beta = 2$ point }
At the collapse point $\beta=2$, the Ursell functions were found
in refs. \cite{Cornu1,Cornu2}.
For fixed fugacity $z$, they read
\begin{subequations}
\begin{eqnarray}
U_{\sigma,\sigma}(r) & = & - \left( {m^2 \over 2\pi} \right)^2 
K_0^2(mr) \label{4.32a} \\
U_{\sigma,-\sigma}(r) & = & \left( {m^2 \over 2\pi} \right)^2 
K_1^2(mr) \label{4.32b}
\end{eqnarray}
\end{subequations}
where $m=2\pi z$, $K_0$ and $K_1$ are modified Bessel functions 
and $\sigma=\pm$.
Consequently,
\begin{equation} \label{4.33}
U_{\sigma\sigma'}(r) \sim - \sigma \sigma' {m^3 \over 8\pi r} 
\exp ( - 2 m r) \quad \quad \quad {\rm as}\ r\to\infty
\end{equation}
in full agreement with (\ref{3.20}) and (\ref{3.22}).

\renewcommand{\theequation}{5.\arabic{equation}}
\setcounter{equation}{0}

\section{Conclusion}
In this paper, we have derived the exact large-distance behavior
of particle correlation functions for the 2D two-component plasma
by using the form-factor theory of the equivalent sine-Gordon model.
The asymptotic decay is always exponential, with a continuously
varying correlation length.
The correlation length is determined by the particle spectrum
of the sine-Gordon model: in the stability regime of the inverse
temperatures $0<\beta<2$, it is equal to the inverse mass $1/m_1$
of the lightest breather $B_1$, and in the collapse region
$2\le\beta\le 4$, where the breathers disappear, it is equal to 
the inverse total mass $1/(2M)$ of the only present particles, 
the soliton-antisoliton pair $(A_+,A_-)$.
On the other hand, the prefactor inverse-power-law function
changes discontinuously at the collapse point $\beta=2$:
it behaves like $1/\sqrt{r}$ in the stability region $0<\beta<2$
[see formula (\ref{3.11})], like $1/r$ at $\beta=2$ (\ref{3.22})
and like $1/r^2$ for $2<\beta<4$ (\ref{3.23}).
The change in the large-distance behavior of correlation functions
from both sides of point $\beta=2$ is a sign of a singularity, which
probably prevents the construction of an analytic $(\beta-2)$ 
expansion of Ursell functions and other finite statistical quantities, 
although all multi-particle Ursell functions are known at 
the collapse point \cite{Cornu2}.

The region $2\le \beta<4$, in which the thermodynamic collapse
occurs, deserves attention.
The Ursell functions calculated in this region are equivalent
to the corresponding Ursell functions of the lattice version of
the 2D TCP, in the continuum limit when the lattice constant $\to 0$,
as was done at $\beta=2$ in refs. \cite{Cornu1,Cornu2}.
The phenomenon of the Kosterlitz-Thouless phase transition
\cite{Kosterlitz} requires the fundamental presence of a maybe small, 
but nonzero, hard core attached to the charged particles.
This is why we do not observe a typical K-T divergence of
the correlation length $\propto \exp(c/\sqrt{4-\beta})$ as
$\beta\to 4^-$.
On the contrary, since the soliton mass $M$ (\ref{3.23b}) goes
to infinity as $\beta\to 4^-$ for fixed $z$, and even for
$z\sim 4-\beta$ as was considered in the usual analysis of 
the renormalization flow close to the K-T transition, the correlation 
length goes to zero when approching $\beta=4$.
In a certain sense, likewise as $\beta=2$ is the collapse point
for the thermodynamics, $\beta=4$ is the ``collapse'' point
for the Ursell correlation functions of our model of pointlike
particles.
In the interval $0<\beta<2$, the introduction of a small hard core
is a slight perturbation which does not change the thermodynamics
and the correlation functions substantially.

A next natural step is to study inhomogeneous situations when 
the 2D TCP is confined by an impermeable wall of dielectric 
constant $\epsilon_W$.
In such situations, the asymptotic decay of pair correlation functions 
is direction-dependent.
In the case of ideal conductor $(\epsilon_W\to \infty)$ and
ideal dielectric $(\epsilon_W=0)$ rectilinear walls, the plasma can be
formulated as a boundary sine-Gordon model with Dirichlet
\cite{Samaj2} and Neumann \cite{Samaj3} boundary conditions,
respectively.
One of first attempts to determine the form factors of such integrable
boundary theories was made in ref. \cite{Hou}.

\section*{Acknowledgments}
We are indebted to K. Chadan for having helped us to understand
the limits of validity of the integral representation (\ref{4.11}).
The stay of L. {\v S}. in LPT Orsay is supported by a NATO fellowship.
A partial support by Grant VEGA 2/7174/20 is acknowledged.

\end{document}